\documentclass[12pt]{article}
\usepackage{a4wide}
\usepackage{amssymb}
\begin{document}
{\renewcommand{\thefootnote}{\fnsymbol{footnote}}
\begin{center}
{\LARGE  Equivalence of models\\ in loop quantum cosmology and group field theory}

\vspace{1.5em}

Bekir Bayta\c{s}, Martin Bojowald and Sean Crowe\\
\vspace{0.5em}
Institute for Gravitation and the Cosmos,\\
The Pennsylvania State
University,\\
104 Davey Lab, University Park, PA 16802, USA\\
\vspace{1.5em}
\end{center}
}

\setcounter{footnote}{0}

\begin{abstract}
 The paradigmatic models often used to highlight cosmological
  features of loop quantum gravity and group field theory are shown to be
  equivalent, in the sense that they are different realizations of the same
  model given by harmonic cosmology. The loop version of harmonic cosmology is
  a canonical realization, while the group-field version is a bosonic
  realization. The existence of a large number of bosonic realizations
  suggests generalizations of models in group field cosmology.
\end{abstract}

\section{Introduction}

Consider a dynamical system given by a real variable, $V$, and a complex
variable, $J$, with Poisson brackets
\begin{equation} \label{VJ}
 \{V,J\}= i\delta J\quad,\quad \{V,\bar{J}\}=-i\delta\bar{J}\quad,\quad
 \{J,\bar{J}\}= 2i\delta V
\end{equation}
for a fixed real $\delta$. If we identify
$H_{\varphi}^{\delta}=\delta^{-1}{\rm Im}J=-i(2\delta)^{-1}(J-\bar{J})$ as a
Hamiltonian generating evolution in some parameter $\varphi$, the equations of
motion are solved by
\begin{eqnarray} \label{Vphi}
 V(\varphi)&=& A\cosh(\delta \varphi)-B\sinh(\delta \varphi)\\
 {\rm Re}J(\varphi)&=& A\sinh(\delta \varphi)-B\cosh(\delta \varphi)\,.
\end{eqnarray}
The brackets (\ref{VJ}) belong to the Lie algebra ${\rm su}(1,1)$ and have the
Casimir $R=V^2-|J|^2$. If $R$ is required to be zero, we obtain
$A^2-B^2-(\delta H_{\varphi}^{\delta})^2=0$ and therefore there is some $\varphi_0$
such that $A/(\delta H_{\varphi}^{\delta})=\cosh(\delta\varphi_0)$ and $B/(\delta
H_{\varphi}^{\delta})=-\sinh(\delta\varphi_0)$. The solution (\ref{Vphi}) then reads
\begin{equation} \label{V}
 V(\varphi)= \delta H_{\varphi}^{\delta}\cosh(\delta(\varphi-\varphi_0))
\end{equation}
and displays the paradigmatic behavior of the volume of a bouncing universe
model. This construction defines harmonic cosmology
\cite{BouncePert,Harmonic}; see also \cite{GroupLQC} for further properties
related to ${\rm su}(1,1)$, in particular group coherent states.

The bouncing behavior can also be inferred from an effective Friedmann
equation that describes modified evolution of the scale factor giving rise to
the volume $V$. To do so, we should provide a physical interpretation to the
time parameter $\varphi$ used so far. A temporal description, shared by some
models of loop quantum cosmology \cite{LivRev,ROPP} and group field cosmology
\cite{GFTCosmo,GFTCosmo2,GFTCosmo3,GFTLattice,GFTPerturb}, is a so-called
internal time \cite{GenHamDyn1}: The parameter $\varphi$ is proportional to
the value of a scalar field $\phi$ as a specific matter contribution devised
such that $\phi$ is in one-to-one correspondence with some time coordinate
such as proper time $\tau$. The scalar $\phi$ itself can then be used as a
global time.  Its dynamics must be such that its momentum $p_{\phi}$ never
becomes zero --- ``time'' $\phi$ then never stops. With a standard isotropic
scalar Hamiltonian
\begin{equation}\label{hphi}
 h_{\phi}=\frac{1}{2}\frac{p_{\phi}^2}{V}+VW(\phi)\,,
\end{equation}
this condition is fulfilled only for vanishing potential $W(\phi)$, such that
$p_{\phi}$ is conserved. The scalar should therefore be massless and without
self-interactions. With these conditions, the conserved momentum $p_{\phi}$
generates ``time'' translations in $\phi$, and can therefore be
identified with the evolution generator $H^{\delta}_{\varphi}$ introduced
above. In order to match with coefficients in the Friedmann equation derived
below, we set 
\begin{equation} \label{pH}
 p_{\phi}= \sqrt{12\pi G} H_{\varphi}^{\delta}\,.
\end{equation}

The Hamiltonian (\ref{hphi}) also allows us to derive a relationship between
$\phi$ and proper time $\tau$, measured by co-moving observers in an isotropic
cosmological model. Proper-time equations of motion are determined by Poisson
brackets with the Hamiltonian constraint, to which (\ref{hphi}) provides the
matter contribution. Therefore,
\begin{equation}
 \frac{{\rm d}\phi}{{\rm d}\tau}= \{\phi,h_{\phi}\}= \frac{p_{\phi}}{V}\,.
\end{equation}
Writing proper-time derivatives with a dot and using $V=a^3$ to introduce the
scale factor $a$, the chain rule then implies
\begin{equation}
\left(\frac{\dot{a}}{a}\right)^2= \left(\frac{\dot{\phi}}{3V}\frac{{\rm
      d}V}{{\rm  d}\phi}\right)^2= \frac{p_{\phi}^2}{9V^4} \left(\frac{{\rm
      d}V}{{\rm d}\phi}\right)^2
\end{equation}
in which
\begin{equation}
 \frac{1}{V^2}\left(\frac{{\rm d}V}{{\rm d}\phi}\right)^2=
 \frac{1}{V^2}\{V,p_{\phi}\}^2= 12\pi G \frac{ ({\rm
     Re}J)^2}{V^2}=
12\pi G\left( 1-\frac{\delta^2p_{\phi}^2}{12\pi G V^2}\right)
\end{equation}
follows from the $\phi$-equations of motion, the zero Casimir $R=0$, and the
identification (\ref{pH}) with $H^{\delta}_{\varphi}=\delta^{-1}{\rm
  Im}J$. Putting everything together,
\begin{equation} \label{EffFried}
\left(\frac{\dot{a}}{a}\right)^2= \frac{4\pi G}{3} \frac{p_{\phi}^2}{V^2}
\left(1-\frac{\delta^2p_{\phi}^2}{12\pi GV^2}\right) = \frac{8\pi G}{3}
\rho_{\phi}\left(1-\frac{\delta^2\rho_{\phi}}{6\pi G}\right) 
\end{equation}
with the energy density $\rho_{\phi}=\frac{1}{2}p_{\phi}^2/a^6$ of the free,
massless scalar. Upon rescaling $\delta=4\pi G\tilde{\delta}$, this effective
Friedmann equation agrees with what has been derived in loop quantum
cosmology, following \cite{AmbigConstr}.

Harmonic cosmology can be obtained as a deformation of a certain model of
classical cosmology. In the limit of vanishing $\delta$,
$H_{\varphi}^0=\lim_{\delta\to0}H_{\varphi}^{\delta}$ has Poisson bracket
\begin{equation}
 \{V,H_{\varphi}^0\}=  \lim_{\delta\to 0}{\rm Re}J\,.
\end{equation}
For finite $H_{\varphi}^0$, we must have $\lim_{\delta\to0}{\rm Im}J=0$, such
that the vanishing Casimir implies $\lim_{\delta\to 0}{\rm Re}J=V$. Therefore, 
\begin{equation}
 \{V,H_{\varphi}^0\}=  V
\end{equation}
with an exponential solution $V(\phi)=\exp(\sqrt{12\pi G}\phi)$ that no longer
exhibits a bounce. Moreover, noticing that
\begin{equation}
 \{V,V^{-1}H_{\varphi}^0\}=1\,,
\end{equation}
we can identify $H_{\varphi}^0/V=P$ with the momentum canonically conjugate to
$V$ in the limit of $\delta\to0$. Therefore, 
\begin{equation} \label{VP}
 H_{\varphi}^0=VP
\end{equation}
is quadratic. Squaring this equation, we find
\begin{equation}
 P^2=\frac{(H_{\varphi}^0)^2}{V^2}= \frac{p_{\phi}^2}{12\pi GV^2}
\end{equation}
which, upon relating $P=\dot{a}/(4\pi G a)$ to the Hubble parameter and $V$
with the scale factor cubed, is equivalent to the Friedmann equation of an
isotropic, spatially flat model sourced by a free, massless scalar field with
momentum $p_{\phi}$:
\begin{equation}
 \left(\frac{\dot{a}}{a}\right)^2= \frac{8\pi G}{3}\rho_{\phi}\,.
\end{equation}

\section{Loop quantum cosmology as a canonical realization of harmonic
  cosmology}

It is of interest to construct a canonical momentum $P$ of $V$ also in the
case of non-zero $\delta$. The pair $(V,P)$ will then be Darboux coordinates
on symplectic leaves of the Poisson manifold defined by (\ref{VJ}), and the
full (real) three-dimensional manifold will have Casimir--Darboux coordinates
$(V,P,R)$. Following the methods of \cite{Bosonize}, we can construct such a
momentum directly from the brackets (\ref{VJ}).

Suppose we already know the momentum $P$. The Poisson bracket of any function
on our manifold with $V$ then equals the negative derivative by $P$. In
particular,
\begin{eqnarray}
 \frac{\partial {\rm Im}J}{\partial P} &=& -\{{\rm Im}J,V\}= \delta  {\rm
   Re}J\\ 
 \frac{\partial {\rm Re}J}{\partial P} &=& -\{{\rm Re}J,V\}= -\delta  {\rm Im}J
\end{eqnarray}
while $\partial V/\partial P=0$. Up to a crucial sign, these equations are
very similar to our equations of motion in the preceding section, and the same
is true for their solutions:
\begin{eqnarray}
 {\rm Im}J(V,P)= A(V)\cos(\delta P)-B(V)\sin(\delta P)\\
 {\rm Re}J(V,P)= -A(V)\sin(\delta P)-B(V)\cos(\delta P)\,.
\end{eqnarray}
Since we are now dealing with partial differential equations, the previous
constants $A$ and $B$ are allowed to depend on $V$.

Given these solutions, we can evaluate the Casimir
\begin{equation}
 R=V^2-|J|^2= V^2-A(V)^2-B(V)^2\,.
\end{equation}
If it equals zero, we have $A(V)^2+B(V)^2=V^2$, and there is a $P_0$ such that
$A(V)/V=- \sin(\delta P_0)$ and $B(V)/V=-\cos(\delta P_0)$. Thus,
\begin{eqnarray}
 {\rm Im}J(V,P)= V\sin(\delta (P-P_0))\\
 {\rm Re}J(V,P)= V\cos(\delta (P-P_0))
\end{eqnarray}
or
\begin{equation}\label{JP}
 J(V,P)=V\exp(i\delta (P-P_0))\,.
\end{equation}
The canonical realization of (\ref{VJ}), given by Casimir--Darboux coordinates
$(V,P,R)$, identifies $J$ as a ``holonomy modification'' of the classical
Hamiltonian (\ref{VP}), in which the Hubble parameter represented by the
momentum $P$ is replaced by a periodic function of $P$. The vanishing Casimir,
$R=0$, then appears as a reality condition for $P$ in (\ref{JP}).

We conclude that the paradigmatic bounce model of loop quantum cosmology,
analyzed numerically in \cite{APS}, is a canonical realization of harmonic
cosmology.

\section{Group field theory as a bosonic realization of harmonic cosmology}

The canonical realization constructed in the preceding section is faithful:
the number of Darboux coordinates agrees with the rank of the Poisson tensor
given by (\ref{VJ}), and the number of Casimir coordinates agrees with the
co-rank. If one drops the condition of faithfulness, inequivalent
realizations can be constructed which even locally are not related to the
original system by a canonical transformations. We will call ``realization
equivalent'' any two systems that are realizations of the same model. This
notion of equivalence therefore generalizes canonical equivalence. As we
will show now, this generalization is crucial in relating loop quantum
cosmology to group field theory.

\subsection{Bosonic realizations}

Instead of canonical realizations, one may consider bosonic realizations,
replacing canonical variables, $(q,p)$ such that $\{q,p\}=1$, with classical
versions of creation and annihilation operators, $(z,\bar{z})$ such that
$\{\bar{z},z\}=i$. The map $z=2^{-1/2}(q+ip)$ defines a bijection between
canonical and bosonic realizations.

The brackets (\ref{VJ}) correspond to the Lie algebra ${\rm su}(1,1)$. A
different real form of this algebra, ${\rm sp}(2,{\mathbb R})$, has a large
number of (non-faithful) bosonic realizations given by the special case of
$N=1$ in the family of realizations
\begin{equation} \label{NonFaithful}
 A^{(n)}_{ij} = \sum_{\alpha=1}^n \bar{z}_{i\alpha}\bar{z}_{j\alpha}\quad,\quad
 B^{(n)}_{ij} = \sum_{\alpha=1}^n z_{i\alpha}z_{j\alpha} \quad,\quad
 C^{(n)}_{ij} = \frac{1}{2}\sum_{\alpha=1}^n \left(\bar{z}_{i\alpha}z_{j\alpha}+
   z_{j\alpha}\bar{z}_{i\alpha}\right)
\end{equation}
of ${\rm sp}(2N,{\mathbb R})$ \cite{Collective,DynCollective,Bosonsp4,BosSymp}
with relations
\begin{eqnarray} \label{ABCRel}
 &&[A_{ij},A_{i'j'}]=0=[B_{ij},B_{i'j'}]\\
 && [B_{ij},A_{i'j'}] = C_{j'j}\delta_{ii'}+ C_{i'j}\delta_{ij'}+
 C_{j'i}\delta_{ji'}+ C_{ii'}\delta_{jj'}\\
 && [C_{ij},A_{i'j'}]= A_{ij'}\delta_{ji'}+A_{ii'}\delta_{jj'}\\
 && [C_{ij},B_{i'j'}]= -B_{jj'}\delta_{ii'}- B_{ji'}\delta_{ij'}\\
 && [C_{ij},C_{i'j'}]= C_{ij'}\delta_{i'j}- C_{i'j}\delta_{ij'}\,.
\end{eqnarray}
The indices take values in the ranges $\alpha=1,\ldots,n$ and
$i,j=1,\ldots,N$, where $i\leq j$ in $A_{ij}$ and $B_{ij}$. There are $2nN$
real degrees of freedom in the bosonic coordinates $z_{i\alpha}$, while ${\rm
  sp}(2N,{\mathbb R})$ has dimension $N(2N+1)$. 

For $N=1$, we have three generators
\begin{equation}  \label{ABC}
 A^{(n)} = \sum_{\alpha=1}^n \bar{z}_{\alpha} \bar{z}_{\alpha}\quad,\quad
 B^{(n)} = \sum_{\alpha=1}^n z_{\alpha}z_{\alpha} \quad,\quad
 C^{(n)} = \frac{1}{2}\sum_{\alpha=1}^n \left(\bar{z}_{\alpha}z_{\alpha}+
   z_{\alpha}\bar{z}_{\alpha}\right)
\end{equation}
with relations
\begin{equation}
 [A^{(n)},B^{(n)}]=C^{(n)}\quad,\quad [A^{(n)},C^{(n)}]=-2A^{(n)}\quad,\quad
 [B^{(n)},C^{(n)}]=2B^{(n)} \,.
\end{equation}
For any $n$, the identification
\begin{equation} \label{ABCJ}
 A^{(n)}=i\bar{J}/\delta\quad,\quad B^{(n)}=iJ/\delta\quad,\quad C^{(n)}=2iV/\delta
\end{equation}
relates these brackets to (\ref{VJ}).

\subsection{Model of group field theory}

In \cite{GFTToy}, a toy model of group field theory has been derived that
produces bouncing cosmological dynamics for the number observable of certain
microscopic degrees of freedom. Starting with a tetrahedron, the model assigns
annihilation and creation operators to the sides, which change the area in
discrete increments. For an isotropic model, the four areas should be
identical, and their minimal non-zero value is determined by a quantum number
$j=1/2$, modelling the discrete nature through a spin system following the
loop paradigm \cite{AreaVol}. Each isotropic excitation has the
``single-particle'' Hilbert space $\left(1/2\right)^{\otimes 4}$ which
contains a unique spin-2 subspace. Since this subspace consists of totally
symmetric products of the individual states, it is preferred by the condition
of isotropy. Restriction to the spin-2 subspace then implies a 5-dimensional
single-particle Hilbert space with complex-valued bosonic variables $A_i$.

A simple non-trivial dynamics is then proposed \cite{GFTToy} by the action
\begin{equation} \label{S}
 S=\int{\rm d}\phi \left(\frac{1}{2}i \left(A_i^*\frac{{\rm d}A^i}{{\rm
         d}\phi}- \frac{{\rm d}A_i^*}{{\rm d}\phi}A^i\right)- {\cal
     H}(A^i,A_j^*)\right)
\end{equation}
in internal time $\phi$. The first term indeed implies bosonic Poisson
brackets $\{A_i^*,A^j\}=i\delta_i^J$. The second term is fixed by proposing a
squeezing Hamiltonian
\begin{equation} \label{HA}
 {\cal H}(A^i,A_j^*)= \frac{1}{2}i \lambda \left(A_i^*A_j^*g^{ij}-
   A^iA^jg_{ij}\right)
\end{equation}
with a coupling constant $\lambda$ and a constant metric $g_{ij}$ with inverse
$g^{ij}$. The metric is defined through an identification of the spin-$2$
index $i$ with all totally symmetric combinations of four indices
$B_I\in\{1,2\}$ taking two values, such that
\begin{equation}
 g_{(B_1B_2B_3B_4)(C_1C_2C_3C_4)}=
 \epsilon_{(B_1(C_1}\epsilon_{B_2C_2} \epsilon_{B_3C_3} \epsilon_{B_4)C_4)}
\end{equation}
with separate total symmetrizations of $\{B_1,B_2,B_3,B_4\}$ and
$\{C_1,C_2,C_3,C_4\}$, respectively, and the usual totally antisymmetric
$\epsilon_{BC}$. Ordering index combinations as 
\begin{equation}
 i\in (1,2,3,4,5)= (1111,(1112),(1122),(1222),(2222))\,,
\end{equation}
the metric can be determined explicitly as the matrix
\begin{equation} \label{metric}
 g = \left(\begin{array}{ccccc} 0&0&0&0&1\\ 0&0&0&-1&0\\ 0&0&1&0&0\\
     0&-1&0&0&0\\ 1&0&0&0&0\end{array}\right)\,.
\end{equation}

A second crucial observable, in addition to the Hamiltonian, is the excitation
number,
\begin{equation} \label{VA}
 V=\frac{1}{2}\left(A_i^*A^i+ A^iA_i^*\right)\,,
\end{equation}
identified with the cosmological volume following group field cosmology. This
volume evolves in internal time $\phi$ according to the Hamiltonian
${\cal H}$. Solutions for $V(\phi)$, derived in \cite{GFTToy}, show bouncing
behavior (\ref{V}) that can be modeled by the effective Friedmann equation
(\ref{EffFried}). 

We can now readily show that this behavior is not a coincidence: The metric
(\ref{metric}) has eigenvalues $+1$ with three-fold degeneracy and $-1$ with
two-fold degeneracy.  Diagonalizing it by an orthogonal matrix gives
linear combinations $z_{\alpha}$ of the $A^i$ and $A_i^*$ that preserve the
bosonic bracket $\{A_i^*,A^j\}=i\delta_i^j$, defining a bosonic
transformation:
\begin{equation}
 z_1=\frac{1}{\sqrt{2}}(A^1+A^5)\quad,\quad
 z_2=\frac{1}{\sqrt{2}}(A^2-A^4)\quad,\quad Z_3=A^3
\end{equation}
for eigenvalue $+1$, and
\begin{equation}
 z_4=\frac{1}{\sqrt{2}}(A^1-A^5)\quad,\quad
 z_5=\frac{1}{\sqrt{2}}(A^2+A^4)
\end{equation}
for eigenvalue $-1$.

We can deal with the negative eigenvalues in two ways. First, multiplication
of $z_4$ and $z_5$ with $i$ preserves the bosonic bracket and leads to a
metric $g'_{ij}=\delta_{ij}$. We then have ${\cal
  H}=\frac{1}{2}i\lambda(A^{(5)}-B^{(5)})$ for (\ref{HA}) and $V=C^{(5)}$ for
(\ref{VA}).  Alternatively, using only diagonalization by an orthogonal
matrix, we have
\begin{equation}
 {\cal H} = \frac{1}{2}i\lambda\left(A^{(3)}-B^{(3)}- (A^{(2)}-B^{(2)})\right)
\end{equation}
and
\begin{equation}
 V=C^{(3)}+C^{(2)}
\end{equation}
where $z_1$, $z_2$ and $z_3$ contribute to the $n=3$ realization, and $z_4$
and $z_5$ to $n=2$.  Observing (\ref{ABCJ}) and the fact that the relations
(\ref{VJ}) are invariant under changing the sign of $J$, the volumes and
Hamiltonians in both loop quantum cosmology and group field theory are
identified with the same generators in harmonic cosmology. The models of loop
quantum cosmology and group field theory are therefore realization equivalent.

\section{Implications and further directions}

There is an immediate application of our result to the appearance of
singularities in the model \cite{GFTToy} of group field cosmology. As argued
in this paper, because the volume is derived from the positive number operator
of microscopic excitations $A^i$, it can be zero only at a local minimum,
which requires $V(\phi_{\rm min})=0$ and ${\rm d}V/{\rm d}\phi=0$ at some
internal time $\phi_{\rm min}$. The combination of these two conditions is
quite restrictive, and \cite{GFTToy} concludes that a singularity (zero
volume) can be reached only for a small number of initial conditions.

However, our identification of the model of \cite{GFTToy} as a bosonic
realization of harmonic cosmology suggests a more cautious approach to the
singularity problem. In ${\rm su}(1,1)$, there is no positivity condition on
the generator that corresponds to the volume $V$. The bosonic realization in
terms of microscopic excitations $A^i$ is therefore local, in the sense that
the $A^i$ are local coordinates on the Poisson manifold that realizes harmonic
cosmology, and $V=0$ is at the boundary of a local chart. Accompanying
$V(\phi_{\rm min})=0$ with ${\rm d}V/{\rm d}\phi=0$ is therefore unjustified
unless one can show that evolution never leaves a local chart. The condition
$V(\phi_{\rm min})=0$ is not as restrictive as the combination, and it leaves
more room for solutions that reach zero volume. (These solutions may still be
considered non-singular if there is a unique Hamiltonian that evolves
solutions through zero volume. In loop quantum cosmology, evolving through
$V=0$ is interpreted as changing the orientation of space
\cite{Sing,IsoCosmo}.)

In harmonic cosmology, further generalizations of the model used here have
already been explored in some detail. The new relationship with group field
theory suggests similar generalizations also on the group-field side of the
equivalence. For instance, harmonic cosmology can be defined for any power-law
$Q=a^p$ replacing $V=a^3$, describing a quantization ambiguity that
corresponds to lattice refinement of an underlying discrete geometry
\cite{InhomLattice,CosConst}. The same algebra, with arbitrary exponent $p$,
can then be realized bosonically, suggesting related group-field
models. (While the power-law $V=a^3$ is preferred at large volume because it
avoids an expansion of the discrete scale to macroscopic size, a different
power low may well be relevant near a spacelike singularity.)

Another parameter related to the relation $V=a^3$ is the averaging volume
$V_0$ used to define the isotropic model. We have implicitly assumed $V_0=1$
in order to focus on algebraic properties; in general, we have $V=V_0a^3$
where $V_0$ is computed as the coordinate volume of the averaging
region. Classical equations do not depend on $V_0$, but quantum corrections
do, as can be seen here from the fact that in the action (\ref{S}) the
Hamiltonian ${\cal H}$ is proportional to $V_0$, but the symplectic term is
not. The microcopic action is then not invariant under changing
$V_0$. Implications of a relation between $V_0$ and the infrared scale of an
underlying field theory \cite{MiniSup} are of importance for the
interpretation of quantum cosmology \cite{Infrared}, and similar conclusions
should hold true in group-field cosmology.

In classical harmonic cosmology, the Casimir $R=0$ is exactly zero, but this
value usually changes in the presence of quantum corrections
\cite{BouncePert,Harmonic, BounceSqueezed}. The bouncing behavior (\ref{Vphi})
is no longer guaranteed if $R<0$ and $|R|>(\delta H_{\varphi}^{\delta})^2$,
because $V(\phi)$ behaves like a sinh under these conditions. These conditions
require large quantum corrections, greater than the matter density related to
$p_{\phi}^2$. They are therefore unlikely to be fulfilled in a macroscopic
universe. However, as pointed out in \cite{Infrared}, an appeal to the BKL
scenario \cite{BKL} near a spacelike singularity shows that a homogeneous
model is a good approximation only if it has small co-moving volume, given by
the averaging volume $V_0$ mentioned above. Such a tiny region does not
contain much matter energy, which can then easily be surpassed by quantum
corrections in a high-curvature regime: $p_{\phi}\propto V_0$ is suppressed
for small $V_0$, while volume fluctuations $\Delta V$ are not proportional to
$V_0$ because they are bounded from below by the $V_0$-independent $\hbar$ in
uncertainty relations. The genericness of bouncing solutions in loop quantum
cosmology or group-field cosmology is then not guaranteed.

Finally, a large class of microscopic models can be constructed from the
bosonic realizations of harmonic cosmology with arbitrary $n$ in
(\ref{ABC}). The question of whether these are related to group field
cosmology in some way appears to be of interest.

\section*{Acknowledgements}

This work was supported in part by NSF grant PHY-1607414.


\end{document}